\documentclass{aa}

\usepackage[dvips]{graphics} 
\usepackage{epsf} 
\usepackage{amssymb} 
 
\newcommand{\IZ}{IZw~36} 
\newcommand{\IZW}{IZw~18} 
\newcommand{\SBS}{SBS~0335-052} 
\newcommand{\Mark}{Markarian~59} 
\newcommand{\1}{\,{\sc i}} 
\newcommand{\2}{\,{\sc ii}} 
\newcommand{\3}{\,{\sc iii}} 
\newcommand{\4}{\,{\sc iv}} 
 
\newcommand{\6}{\,{\sc vi}} 
 
\begin{document} 

   \title{Abundance differences between the neutral 
and the ionized gas of the dwarf galaxy \IZ}

 
 \author{V. Lebouteiller\inst{1}, D. Kunth\inst{1}, J. Lequeux\inst{2}, 
A. Lecavelier~des~Etangs\inst{1},  
J.-M.~D\'esert\inst{1},  G. H\'ebrard\inst{1} 
	\and A.~Vidal-Madjar\inst{1} 
 }

   \offprints{V. Lebouteiller, \email{leboutei@iap.fr}} 
 
   \institute{Institut d'Astrophysique de Paris, CNRS, 98 bis Boulevard  
Arago, F-75014 Paris, France 
         \and 
           LERMA - B\^atiment A, Observatoire de Paris, 61, Avenue  
de l'Observatoire,  F-75014 Paris, France \\ 
             }

   \date{Received 06/02/03; accepted 10/20/03} 
   \abstract{We present a \textit{FUSE}  spectroscopic study  of the nearby 
gas--rich, metal--deficient blue compact dwarf (BCD) galaxy 
 \IZ. Atomic hydrogen and many metal lines are observed in  absorption 
 against the stellar continuum of  young, massive stars embedded in  
the ionized region. Profile fitting of absorption lines  allowed us to determine  abundances and investigate the chemical composition of the  
 neutral gas. This study present strong evidences that nitrogen
is 16$^{+7}_{-8}$ times less abundant in the neutral gas with respect to the ionized gas (all uncertainties are 2$\sigma$).
Similarly, the oxygen abundance estimated using phosphorus as a tracer is lower in the neutral gas by a factor 8$^{+17}_{-5}$. 
We also find that argon is underabundant by a factor 32$^{+8}_{-7}$ and that  log~(Ar\1/O\1)$<$$-$3.0 which is inconsistent with the solar value $-$2.1$\pm$0.1 (Lodders~2003), implying  that argon is likely ionized into Ar\2\ in the neutral medium.

 } 
 
\titlerunning{Abundances in the neutral gas of \IZ} 
\authorrunning{V. Lebouteiller et al.} 
 
   \maketitle 
%
 
\section{Introduction}

Blue compact dwarfs (BCDs) are gas--rich and relatively unevolved objects, as 
suggested by their low metallicity (between $\sim Z_\odot$/50 and $\sim Z_\odot$/3).  
Their star formation history is thought to be dominated by short, intense 
bursts separated by long quiescent periods (Searle \& Sargent~1972). The present starburst episode 
gives an opportunity  to compare the  chemical abundances of the H\2\ region 
where stars recently formed with those of their surrounding extended neutral gas.  One of the issues is to know whether metal abundances as measured from H\2\ regions reflect the true metallicity of star--forming galaxies and whether the neutral phase is less chemically evolved than the ionized phase.
 Such an attempt has been done in the past  using Hubble Space 
Telescope (\textit{HST}), with inconclusive results because of the limited    
wavelength coverage  and the difficulties in analyzing the saturated $\lambda$1302.2~\AA\ O\1\ line (Kunth et al.~1994 and Pettini~\&~Lipman~1995). The Far Ultraviolet Spectroscopic Explorer 
(\textit{FUSE}, Moos~et~al.~2000) gives access  to many transitions  
of critical species such as H\1, N\1, O\1, Si\2, P\2, Ar\1,  or  Fe\2,  
allowing one to revisit the chemical composition in neutral regions and the 
metal enrichment processes  in nearby galaxies.

The fate of the metals newly produced by starbursts in BCDs is not settled yet. Kunth~\&~Sargent~(1986)  
have emitted the hypothesis that the H\2\ region of \IZW\ has enriched  
itself over the timescale of the burst (i.e. few 10$^6$~yr). The H\2\ region could be   self--polluted, resulting in larger abundances than the surrounding neutral medium.
Another  possibility  is that, once released by massive stars, metals  remain in 
a hot phase, being unobservable immediately through optical and UV emission 
lines. They would then cool into molecular droplets which later on settle  
onto the disk  (Tenorio-Tagle~1996). Mixing with the interstellar medium 
(ISM) has to await photodissociation of the droplets by the UV radiation of  
another generation of stars before their material changes the interstellar medium composition.

 The study of various elements in both  ionized and neutral gas in  
star--forming galaxies should help to constrain the enrichment processes and  
identify  parameters that lead to possible abundance differences such as  
depletion on dust grains, ionization state correction and unprocessed (less chemically evolved) gas in  
the line of sight. 

In this paper we  investigate the ISM of \IZ\  through  
absorption lines of species standing in the line of sight of the   
ionizing stellar complex within the BCD.
\IZ\ is a metal--deficient  BCD with Z=Z$\odot$/14 in the H\2\ region (Viallefond~\& Thuan~1983, hereafter V83). 
The H\1\ distribution shows a core--halo structure. The ionizing cluster is associated with the H\1\ core which contains half of the total H\1\ mass (V83).
The H\1\ halo is diffuse and contains several clumps.

 In order to compare the   
chemical composition of the neutral gas with respect to the ionized gas, we will   consider the abundances of nitrogen, oxygen, silicon, argon, phosphorus, and iron. The  
observations are described in Sect.~\ref{sec:obs} and the data analysis is  
explained in Sect.~\ref{sec:data}. The composition of the neutral gas of 
\IZ\ is derived in Sect.~\ref{sec:compo}. In Sect.~\ref{sec:compa}, we 
discuss the  abundance differences between the ionized and the neutral gas.

 
\section{Observations}\label{sec:obs}

\IZ\ was observed with \textit{FUSE} on 2001, January 13 (observation Q2240101) 
with an integration time of 19350 seconds (7 sub-exposures) and on 2000, May 
9 (observation P1072201) with 6290 seconds (3 sub-exposures). Because of the angular extent of 
 the ionizing cluster, the largest aperture LWRS was chosen (30''$\times$30'').  At a distance of 6~Mpc (Table~\ref{tab:char}), 1'' corresponds to 
 30~pc.

\begin{table}[h!] 
\caption{\small{Characteristics of \IZ.}}\label{tab:char} 
\begin{center} 
\begin{tabular}{ccc} 
\hline 
\hline 
Parameter & \multicolumn{2}{c}{Value} \\ 
\hline 
$\alpha$(2000), $\delta$(2000) & \multicolumn{2}{c}{12$^{\rm h}$26$^{\rm m}$16$^{\rm s}$, +48$^\circ$29'37''} \\ 
$M_{\rm B}$ & \multicolumn{2}{c}{-14.07} \\ 
$v^a$ (km s$^{-1}$)  & \multicolumn{2}{c}{281$\pm$4} \\ 
$D^b$ (Mpc) & \multicolumn{2}{c}{5.8 $<$ $D$ $<$ 7.9  }\\ 
$Z^c$ & \multicolumn{2}{c}{$\approx$ Z$_\odot$/14} \\ 
\hline 
\end{tabular} 
\end{center} 
\begin{list}{}{} 
\item[$^{\mathrm{a}}$] Heliocentric radial velocity. 
\item[$^{\mathrm{b}}$] Distance (Schulte-Ladbeck~et~al.~2001). 
\item[$^{\mathrm{c}}$] Metallicity (V83). 
\end{list} 
\end{table}

Data were recorded through the two LiF channels ($\approx$1000--1200~\AA) and the two SiC channels ($\approx$900--1100~\AA) and processed by the pipeline \texttt{Calfuse 2.2.0}. 
The  spectrum of each channel results from  the co-addition of  
several individual sub-exposures. 
The typical signal-to-noise ratio (S/N)  per resolution element of the  
co-added spectra ranges from 2 below 1000~\AA\ up to 4 above 1000~\AA. 
The final spectrum shows two absorption lines systems separated by  
approximately 1~\AA. The first at $\approx$$-$50~km~s$^{-1}$ is due to the local ISM  of the Milky Way. The second is at 270$^{+5}_{-6}$~km~s$^{-1}$, which  matches  with the heliocentric velocity of \IZ\ (Table~\ref{tab:char}).

\section{Data analysis}\label{sec:data}

To analyze the data, we used the profile fitting procedure \texttt{Owens} developed at the 
Institut  
d'Astrophysique de Paris by Martin Lemoine and the \textit{FUSE} French 
team.  
This program returns the most likely  
values of many free parameters such as  temperatures,  Doppler widths,   
velocities, or  column  
densities through a minimization of the  
difference between the observed and computed profiles of absorption lines. 
Furthermore, the version we used allows changes of the background level, the  
continuum, and the line broadening. The continua were fitted 
by  
zero- to fourth-order polynomials, depending on the spectral region. All the  
coefficients of the polynomials were free. An example of fitted profiles can 
be  
seen in Fig.~\ref{fig:plot}.

We consider independently several groups of species, each group being defined by its turbulent velocity, its temperature, and its heliocentric velocity. 
As shown in Table~\ref{tab:groups},  one group defines  species supposed  
to be mainly present in the neutral gas (see also 
Sect.~\ref{sec:abundances}).  
The two other  groups contain respectively species of the ionized phase and 
 the hot gas (giving  
the absorption lines of O\6).  Molecular hydrogen is discussed separately in 
Sect.~\ref{sec:CD}.

\begin{table*} 
\caption{\small{List of the lines detected in  
\IZ. MW is for the Milky Way, $\lambda$ is the rest wavelength. Velocities of 
each 
 species have been computed  
independently; 
 typical errors are $\pm$10km~s$^{-1}$. Species are distributed into three 
groups (see text). $b$ and $v$ are calculated using line profiles of all 
 species within a group.}}\label{tab:groups} 
\begin{center} 
\begin{tabular}{llrl} 
\hline 
\hline 
Species & Velocity   & $\lambda$ &  comment \\ 
        &  (km s$^{-1}$) & (\AA) & \\ 
\\ 
\hline 
\multicolumn{4}{c}{First group $\equiv$ neutral region of \IZ. $b$=12.2$^{+3.4}_{-4.2}$~km s$^{-1}$, $v$=270$^{+5}_{-6}$~km s$^{-1}$}\\ 
\hline 
H\1\ & & 1025.72 & Ly$\beta$, presence of damping wings (Fig.~\ref{fig:plot}a)\\ 
& & 937.80, 930.75, 926.23, 923.15 & resp. Ly$\epsilon$, Ly$\zeta$, Ly$\eta$, Ly$\theta$\\ 
 
C\2\ & 259 & 1036.34 & strongly saturated, blended with  C\2* from MW and  O\6\ from MW \\ 
 
N\1\ &265 & 1134.98& not saturated (Fig.~\ref{fig:compaNI})\\ 
& & 963.99 & not saturated, blended with N\1\ from MW  \\ 
& & & and P\2\ from \IZ \\ 
& & 954.10, 953.97, 953.65, 953.42  & not saturated, blended with each other\\

O\1\ & 272 & 1039.23 & strongly saturated \\ 
& & 976.45 & saturated, blended with C\3\ from MW \\ 
& & 950.89 & saturated, blended with P\4\ from \IZ \\

Si\2\ & 261 & 1020.70 &  slightly saturated \\ 
 
P\2\ & 270 & 1152.82 &  detected \\
     &    &  963.83 & blended with  N\1\ from MW and N\1\ from \IZ \\

Ar\1\ & & 1048.22 & not detected (Fig.~\ref{fig:ari})\\ 
 
Fe\2\ & 270  & 1144.94 &  not saturated (Fig.~\ref{fig:plot}b) \\ 
& &  1143.23, 1142.37, 1125.45 & not saturated  \\ 
& &  1121.97, 1096.88, 1063.18 & not saturated \\ 
 
\\ 
\hline 
\multicolumn{4}{c}{Second group:~species with ionization potentials$>$29~eV. $b$=24.3$^{+15.9}_{-16.3}$~km s$^{-1}$, $v$=260$^{+8}_{-12}$~km s$^{-1}$}\\ 
\hline 
C\3\ & 261& 977.02 & detected \\ 
N\2\ & 254& 1083.99 & barely detected\\ 
P\4\ & 254 & 950.66 & blended with O\1\ from \IZ \\ 
S\3\ &248 & 1012.50 & detected \\ 
Fe\3\ &264 & 1122.53 & detected \\ 
\\ 
\hline 
\multicolumn{4}{c}{Third group $\equiv$ hot gas. $b$=51.4$^{+35.9}_{-18.7}$~km s$^{-1}$, $v$=235$^{+23}_{-16}$~km s$^{-1}$}\\ 
\hline 
O\6\ & 235 & 1037.62, 1031.93 & detected \\ 
 
\hline 
\end{tabular} 
\end{center} 
\end{table*}

\subsection{The line broadening}\label{sec:lsf}

Taking the observed total line broadening $\sigma_{tot}$ as a free parameter, 
 we used \texttt{Owens} to find its most likely value in \IZ\  spectra, supposing that $\sigma_{tot}$   is constant in the \textit{FUSE} spectral range. We find $\sigma_{tot}$=27$\pm$3~pixels ($\approx$50~km~s$^{-1}$ at $\lambda=1000$~\AA).

The total broadening of the unsaturated absorption lines has several origins:\\
$-$ The instrumental line spread function. We could not use the H$_2$ lines from \IZ\ and the Galactic component, which usually allow  a good  estimation of the instrumental broadening,  since they are not detected in our spectra (Sect.~\ref{sec:molecular}). For a bright point--like source, the  full width at half maximum (FWHM) is $\sigma_{inst}=$11~pixels   (H\'ebrard~et~al.~2002).\\
$-$ The misalignments of individual sub-exposures. The final spectrum of each detection channel is obtained by co-additions  
requiring wavelength shifts of individual sub-exposures. This unavoidably  
introduces some misalignments  because of the low S/N ratio of  each exposure.  We estimate that such  misalignments cause an additionnal broadening of the  LSF of $\sigma_{add} \approx$10~pixels.\\ 
$-$~The spatial distribution of the UV bright stars within the slit. \textit{HST}/FOC images  
(Deharveng et al.~1994) reveal that the main  
concentration has an extent of $\approx$4'', corresponding to a wavelength  
smearing of $\sigma_{spat} \approx$7~pixels.\\ 
$-$~The main source of line broadening is the velocity  
distribution of the absorbing clouds standing in the multiple lines of sight,  
whose width is \textit{a priori} unknown but can be deduced for an unsaturated line by  
$\sigma_{clouds}$=$(\sigma_{tot}^2-\sigma_{inst}^2-\sigma_{spat}^2)^{1/2}$=22~pixels i.e. 40~km~s$^{-1}$.

The influence of the resulting total broadening $\sigma_{tot}$ on the determination of column densities is discussed in  
Sect.~\ref{sec:CD}.

\subsection{The turbulent velocity} 
The turbulent velocity  parameter \textit{b} of the neutral gas is well 
 constrained by  both saturated--profile and Doppler--profile lines.  
From profile fitting we obtain   \textit{b}=12.2$^{+3.4}_{-4.2}$~km~s$^{-1}$ (2$\sigma$ error bars),  
corresponding to a velocity 
 dispersion FWHM of 
 $\sigma_{neutral}$=$2~\sqrt{\rm{ln}~2}~b$=20.3~km~s$^{-1}$.  
This result is consistent with the velocity dispersion measured from 21~cm 
 radio  
observations ($\sigma_{21~cm} \approx$45~km~s$^{-1}$, V83) and the  
derived velocity distribution of the absorbers (see previous section):~$\sigma_{21~cm}^2 \approx \sigma_{neutral}^2 + \sigma_{clouds}^2$. 
 
Of course,  a significant fraction of the gas whose turbulent velocity is  
measured by radio  
observations is behind the ionizing stars in front of which absorption lines 
 from  neutral elements originate, hence the velocity structure in 
our case could be different (Pettini~\&~Lipman~1995).  
The effect of the determination of the turbulent velocity on column  
densities is discussed in Sect.~\ref{sec:CDni}.

\section{Composition of the neutral gas}\label{sec:compo}

\subsection{Column densities}\label{sec:CD} 
 
Column densities are given in Table~\ref{tab:CD}.
Althrough the paper, error bars are 2$\sigma$.
 The multiple lines of sight to the  massive stars that 
contribute to the UV continuum may cross clouds  with different chemical  
composition, so that the reported column densities represent global values  
among all the clouds.

\begin{table}[h!] 
\caption{\small{Column densities derived in \IZ. Errors are given at 2$\sigma$, upper limits at 3$\sigma$. Upper limits for H$_2$ are calculated for the same turbulent velocity as the neutral phase (12.2~km~s$^{-1}$). The Fe\3\ column density is calculated from the Fe\3\ $\lambda$1122.53 line which could  originate from both the ISM and stellar atmospheres.}}\label{tab:CD} 
\begin{center} 
\begin{tabular}{llcc} 
\hline 
\hline 
Species   & log~$N$(cm$^{-2}$) & log~(X/H) & $[$X/H$]^a$\\ 
\hline 
H\1 & $21.30 ^{+0.09} _{-0.10}$ & & \\ 
C\2  & $18.31 ^{+0.18} _{-0.19}$  & & \\ 
C\3  & $14.38 ^{+0.18} _{-0.19}$  &&\\ 
N\1  & $14.42 ^{+0.26} _{-0.22}$ & $-6.88^{+0.36} _{-0.31}$& $-2.71^{+0.47} _{-0.42}$ \\ 
N\2  & $14.02 ^{+1.15} _{-0.98}$ & &\\ 
O\1  & $16.80 ^{+1.85} _{-1.01}$ & $-4.50^{+1.96} _{-1.10}$ & $-1.19^{+2.00} _{-1.15}$\\ 
O\6  & $13.97 ^{+0.16} _{-0.22}$ &&\\ 
Si\2  & $15.34 ^{+0.85} _{-0.28}$ & $-5.96^{+0.96} _{-0.37}$& $-1.50^{+0.97} _{-0.39}$ \\ 
P\2 & $13.00 ^{+0.28} _{-0.36}$ & $-8.30^{+0.36} _{-0.45}$ & $-1.76^{+0.41} _{-0.51}$ \\ 
S\3  & $14.84 ^{+0.43} _{-0.15}$ & &\\ 
Ar\1  & $<13.30$ & $<$$-$8.00 & $<$$-2.55$ \\ 
Fe\2  & $14.51 ^{+0.21} _{-0.14}$ & $-6.79^{+0.31} _{-0.23}$ & $-2.26^{+0.34} _{-0.26}$ \\ 
Fe\3  & $14.47 ^{+0.18} _{-0.14}$ & & \\ 
\hline 
H$_{2\textrm{ (J=0)}}$ & $<14.41$ &&\\ 
H$_{2\textrm{ (J=1)}}$ & $<14.61$&&\\ 
H$_{2\textrm{ (J=2)}}$ & $<14.32$ &²\\ 
\hline 
\end{tabular} 
\end{center} 
\begin{list}{}{}
\item[$^{\mathrm{a}}$]  $[$X/Y$]$=log~(X/Y)$-$log~(X/Y)$_\odot$, where log~(X/Y)$_\odot$ is  the 
solar value. We use the recommended solar values of Lodders~(2003).
\end{list} 
\end{table}

\begin{figure}[t!] 
\hspace{0.4cm} 
\epsfxsize=5.5cm 
\rotatebox{-90}{\epsfbox{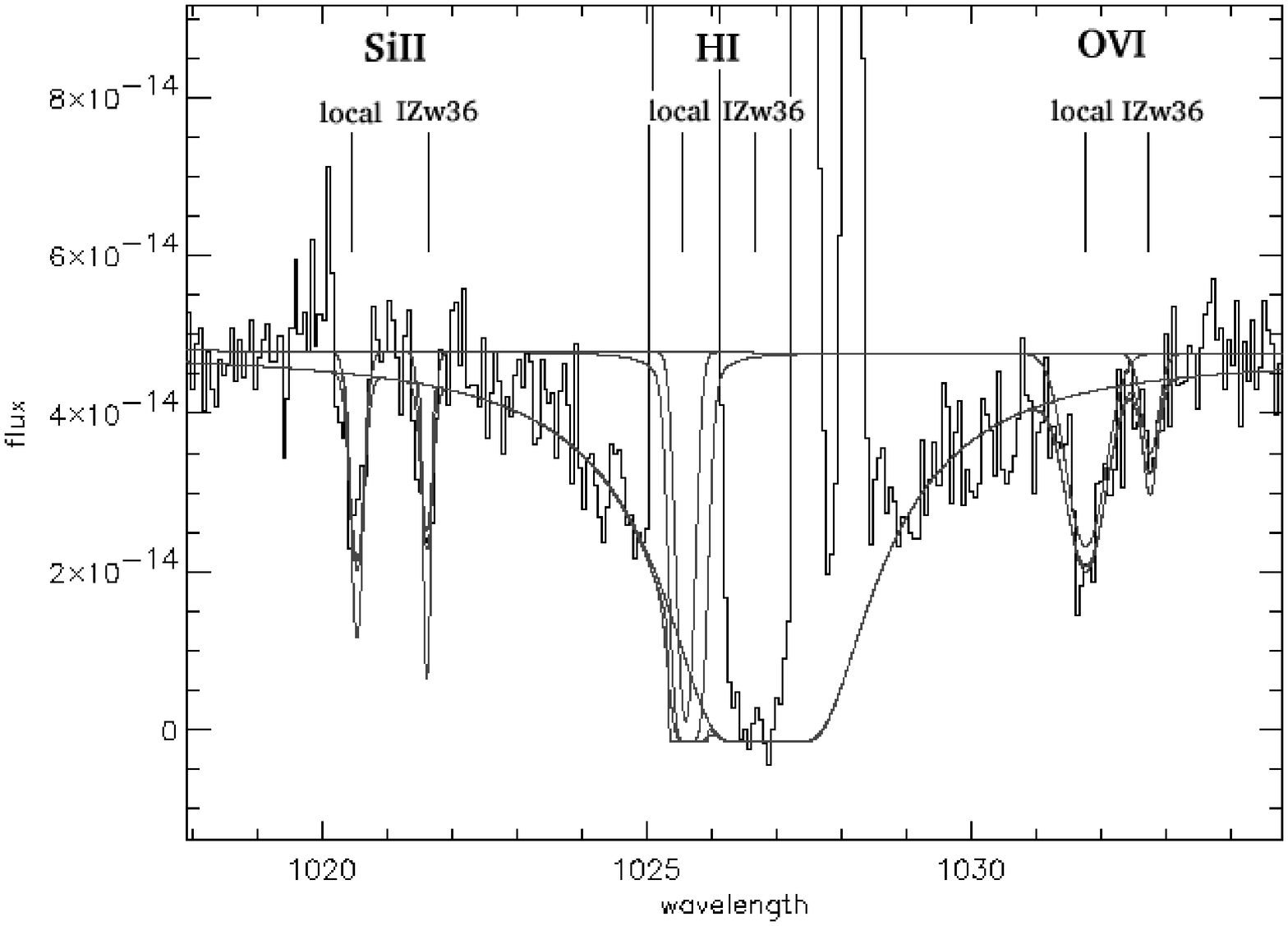}} 
     
\hspace{0.4cm} 
\epsfxsize=5.5cm 
\rotatebox{-90}{\epsfbox{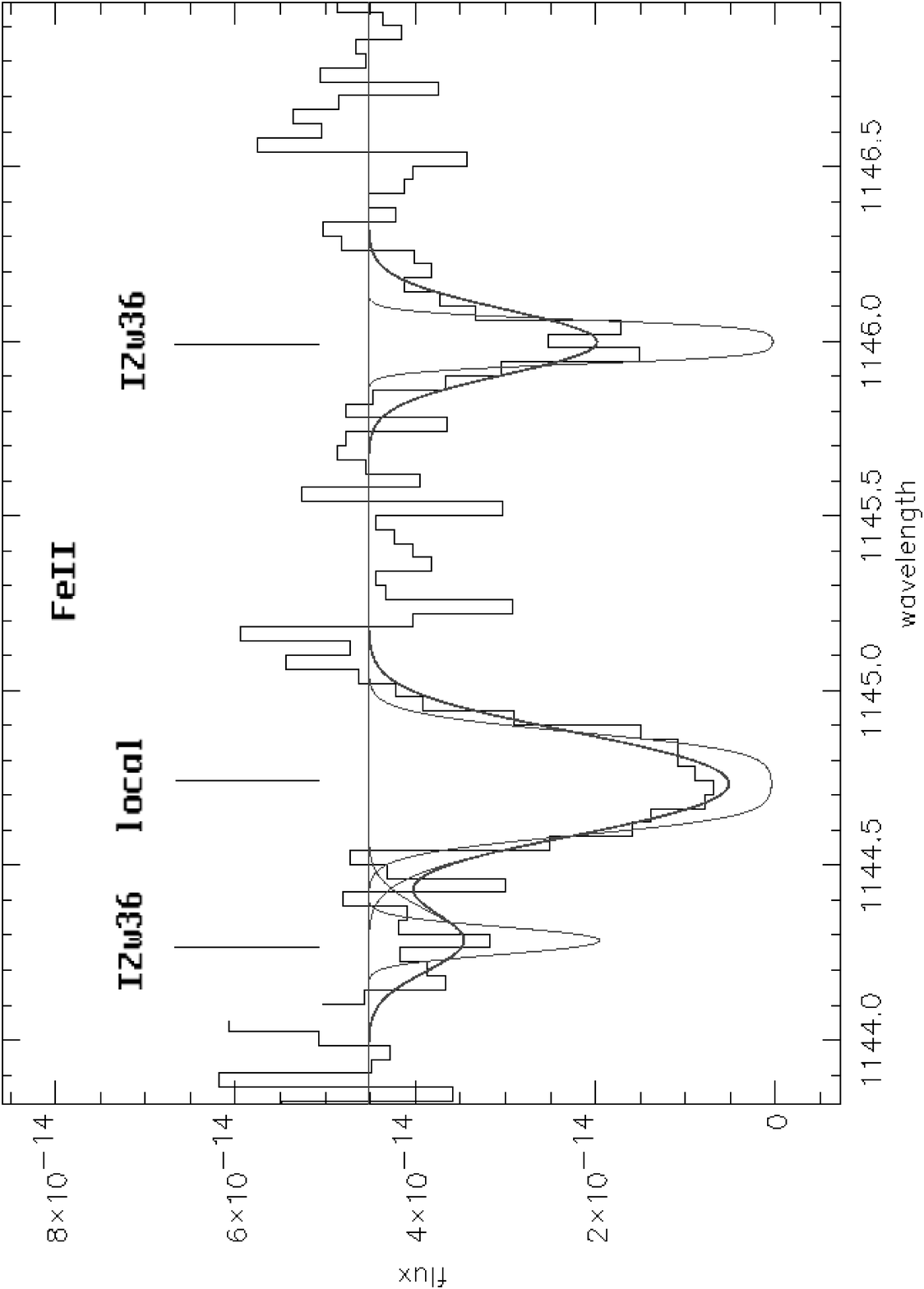}} 
         \caption[]{\small{H\1\ and Fe\2 lines in \IZ. The flux is in  
erg~s$^{-1}$~cm$^{-2}$~\AA$^{-1}$ and the observed wavelength in \AA. 
The data  (histogram)  are binned by a factor 3 for  display purposes.
The thin line  
represents the line profile before the convolution with the total line broadening represented by $\sigma_{tot}$ in Sect.~\ref{sec:lsf} and the  
thick line is the convolved profile. 
For each of the spectral windows, we used the two observations and all the available \textit{FUSE} detection channels (Sect.~\ref{sec:obs}) to perform the fit. Here we show the spectrum of segment LiF1 for the upper plot and of segment LiF2 for the lower plot.
\textbf{a)} \textit{upper} Plot of the  
Ly$\beta$ line. The center of the line is blended with terrestrial  
airglow emission lines. Atomic hydrogen in the Milky Way  is invisible  
because of its low column density $\sim$10$^{19}$~cm$^{-2}$ (estimated from  
Heiles~1975). The Si\2\ $\lambda$1020.70  lines and  O\6\ $\lambda$1031.93 
lines  
from the Milky Way  and \IZ\ are also visible in this spectrum. \textbf{b)}  
\textit{lower} Plot of the strong Fe\2\ $\lambda$1144.94 line.}} 
         \label{fig:plot} 
\vspace{0.1cm} 
\end{figure}

Errors on the column densities essentially depend on where the lines are located on the 
curve  
of growth. In  
particular, saturated lines can give column densities spanning 3 orders of  
magnitude  
when \textit{b} is not  constrained by lines of other species  
standing in the same gas phase.  
Errors  are computed  using the $\Delta\chi^2$ method:~$\chi^2$ is calculated 
as a function of the column density. We use the difference  
$\Delta\chi^2$=$\chi^2-\chi^2_{min}$, where $\chi^2_{min}$ is the best 
 fit to derive the  error bars  (2$\sigma$ error bars correspond to  
$\Delta\chi^2$=4). This method is also  
used to evaluate error bars for $b$ and the  line broadening. Figure~\ref{fig:bdep}  shows an example of a $\Delta\chi^2$ curve.

The determination of the H\1\ column density is not affected by the line broadening $\sigma_{tot}$. However, we have tested several values of $\sigma_{tot}$  in order to test its impact  
 on the  Fe\2, Si\2, P\2, and N\1\ column densities.
 It can be seen in  
Table~\ref{tab:lsf} that for $\sigma_{tot}$  between 12 and 27~pixels,  
 column densities do not vary by more than  a factor 2.5 in the worst  
case (Si\2). A bad determination of the line broadening would not change the
results of Sect.~\ref{sec:compa}.

\begin{table}[t!] 
\caption{\small{Effects of the total line broadening $\sigma_{tot}$ (in pixels, see Sect.~\ref{sec:lsf}) on column densities (log~$N$(cm$^{-2}$)). Note that if $\sigma_{tot}$ is overestimated   
so are the column densities.}}\label{tab:lsf} 
\begin{center} 
\begin{tabular}{lccccc} 
\hline 
\hline 
  & $\sigma_{tot}$=12 & 15 &  21 & 24 & 27 \\ 
\hline 
log~$N$(N\1) & 14.30 & 14.30 &  14.32 & 14.36 & 14.42 \\ 
log~$N$(Si\2) & 15.08 & 15.08 &  15.15 & 15.20 & 15.34 \\ 
log~$N$(P\2) & 12.99 & 13.02 &  13.14 & 13.16 & 13.17 \\
log~$N$(Fe\2) & 14.20 & 14.20 &  14.26 & 14.30 & 14.49 \\ 
\hline 
\end{tabular} 
\end{center} 
\end{table}

\subsubsection{Molecular hydrogen}\label{sec:molecular}

No H$_2$ lines were detected at the redshift of \IZ\    
(Table~\ref{tab:CD}), as for \IZW\ (Vidal-Madjar~et~al.~2000), \Mark\ (Thuan~et~al.~2002), and \SBS\ (Thuan~et~al.~2003). There are several possible  physical 
 explanations for this absence as pointed out by Vidal-Madjar~et~al.~(2000):~the ionizing 
 flux of the massive stars which can be high enough to destroy H$_2$ molecules, the scarcity of dust 
 grains on which H$_2$ molecules are synthetized, and a low H\1\ volumic density. Also, Vidal-Madjar~et~al.~(2000) argue that the molecular medium could be clumpy so that  the few clumps in the lines of sight of the many blue stars in the entrance aperture would not be detected.

\subsubsection{H\1}\label{sec:CDhi}

The H\1\ column density has been calculated using all the available Lyman lines (Ly$\beta$, Ly$\epsilon$, Ly$\zeta$, Ly$\eta$, and Ly$\theta$) but is mainly constrained by the damping wings of the  Ly$\beta$ line (Fig.~\ref{fig:plot}a) since the other H\1\ lines are strongly saturated.
We used a 30~\AA\ wide  window centered on Ly$\beta$ to constrain the continuum by a 4-order polynomial. We find log~$N$(H\1)= $21.30 ^{+0.09} _{-0.10}$. The errors on log~$N$(H\1)  are the smallest amongst the specie we investigate. This is due to  the damping profile of the Ly$\beta$ line which is observed through both LiF and SiC channels.

Our value is somewhat higher than log~$N$(H\1)=21.15 found by V83 who used only the red damping wing of the Ly$\alpha$ line from \textit{IUE} observations,  but is certainly better constrained in our case because of the two damping wings of the Ly$\beta$ line and because of the better resolution and the better signal-to-noise ratio of our observations.

\subsubsection{N\1\ and N\2}\label{sec:CDni}

\begin{figure}[h!] 
\hspace{0.2cm} 
\epsfxsize=8.0cm 
\rotatebox{180}{\epsfbox{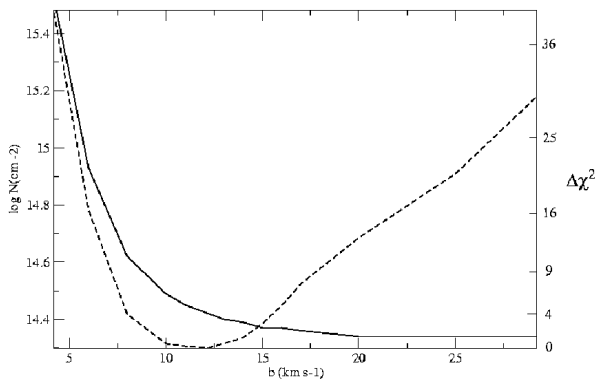}} 
      \caption[]{\small{Effects of the determination of the turbulent velocity \textbf{b} on the N\1\ column density (solid line). Errors in \textit{b} do not affect much the  column density determination. The $\Delta\chi^2$ curve (dashed line) is relatively narrow and  is minimal for $b$=12.2~km~s$^{-1}$. At 2$\sigma$ ($\Delta\chi^2$=4) we obtain $b$=12.2$^{+3.4}_{-4.2}$ corresponding to log~$N$(N\1)=14.43$^{+0.19}_{-0.06}$.}} 
         \label{fig:bdep} 
\vspace{0.2cm} 
\end{figure}

N\1\ is the most useful species in our study since most of it must be  present in the  
neutral ISM. Six N\1\ lines were used (Table~\ref{tab:groups}).
These lines are not saturated so that the determination of log~$N$(N\1) is not strongly  
dependent on the turbulent velocity \textit{b} (Fig.~\ref{fig:bdep}). 
We obtain log~$N$(N\1)=$14.42 ^{+0.26} _{-0.22}$.

The $\lambda$1083.99~N\2\ line is barely detected since it is located in a spectral region  which is very noisy. We are however able to calculate a column density:~log~$N$(N\2)=$14.02 ^{+1.15} _{-0.98}$.

\subsubsection{O\1\ and P\2}\label{sec:CDoi}
 
The three O\1\ lines in \IZ\  are either blended or 
strongly saturated (Table~\ref{tab:groups}), resulting in large errors (Table~\ref{tab:CD}). 
 
The $\lambda$963.83~P\2\ line has a  large oscillator strength ($f$=1.25)  
and is detectable even for  low P\2\ column densities. In \IZ\ spectra,  
this line is blended with N\1\ lines from the Milky Way and from \IZ. Given the fact that the N\1\ column density is  
well determined and that the P\2\ line is not saturated, the blending can  
be corrected for.
We also use the $\lambda$1152.82 P\2\ line with a fainter oscillator strength to constrain the   
final P\2\ column density  as
log~$N$(P\2)$=13.00^{+0.28}_{-0.36}$.

\subsubsection{Si\2, Ar\1, Fe\2}\label{sec:CDothers}

The Si\2\ column density is only given by the $\lambda$1020.70 line which is slighly saturated. Unfortunately, the other strong Si\2\ line at 989.90~\AA\ is contaminated by terrestrial airglows.
 
\begin{figure}[t!] 
\hspace{0.4cm} 
\epsfxsize=5.5cm 
\rotatebox{-90}{\epsfbox{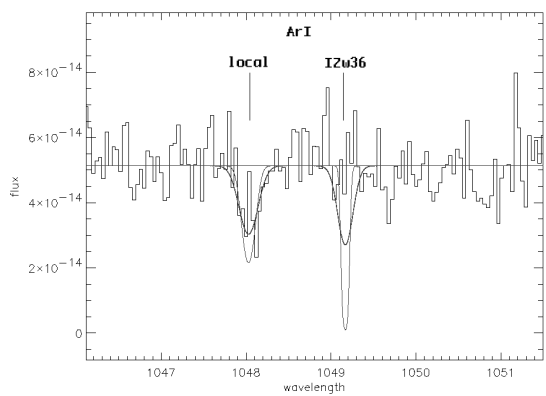}}  
 
\caption[]{\small{Lack of argon in \IZ.  See Fig.~\ref{fig:plot} for a description of the plot. Data are binned by a factor 2 for  display purposes. The line plotted at the redshift of \IZ\ is for log~$N$(Ar\1)=14.0 and \textit{b}=12.0~km~s$^{-1}$.}} 
  \label{fig:ari} 
\vspace{0.1cm} 
\end{figure}

The other  species lead to reliable column densities: Ar\1\ (Fig.~\ref{fig:ari}) for which we have a good upper limit, and Fe\2\ (Fig.~\ref{fig:plot}b) with a large collection of lines.

\subsection{Abundances in the neutral region}\label{sec:abundances} 
In order to derive total abundances from ionic column densities,  
it is necessary to know  the ionization states of each species in the neutral 
 region. 
 
In principle, in the H\1\ region we expect to find all  elements with an  
ionization potential larger than that of hydrogen (13.6~eV) as neutral atoms. 
 Elements with  smaller ionization potentials must be found as  
single--charged ions. For example, we can safely assume that iron  
and silicon exist mainly as  Fe\2\ and Si\2\ in the neutral region  and  
that the Fe\2\ and Si\2\ lines are essentially produced there. We have 
 checked that the column densities of these ions are considerably smaller  
in the H\2\ region using the model E2E1 from the grid of  Stasi\'nska (1990).
The model E2E1 agrees the best with the physical parameters of the H\2\ region derived by V83.
 The resulting ionic fractions of  Si\2\ and Fe\2\ in the H\2\ region in the 
 best fitting model are only a few percents.

The situation is however not so simple for elements with an ionization 
 potential not much larger than that of hydrogen, because UV photons able  
to produce their ionization may be present in low--density, partly--ionized 
 regions of the ISM. For example, Ar\1\ (ionization potential 15.8~eV) is  
easy to ionize because  of its  relatively large photoionization cross  
section (Sofia~\&~Jenkins~1998).  
Therefore, in regions with an ionization degree $n_e/n_H$$\approx$0.2, the actual 
Ar/H can be larger by 0.2 to 0.7~dex than Ar\1/H\1, depending on the hardness 
of the far--UV radiation. The situation is similar but less severe for O\1\ 
and N\1\ (ionization potentials 13.6~eV and 14.5~eV respectively). Indeed, 
in the conditions of the Local Cloud for which the correction for Ar\1/H\1\ 
is 0.36~dex, the corresponding correction for N\1/H\1\ is only 0.05~dex 
(Sofia~\&~Jenkins~1998). 
However the ionization correction in BCDs could be different and could reach 0.2-0.3~dex.
We will first neglect this correction, and because little 
N\1\ and O\1\ are expected in the H\2\ region, we assume that the N\1\ and 
O\1\ lines are representative of the total nitrogen and oxygen abundances in the  neutral region.

The abundances of the neutral region can now be compared with the ionized region  values. For the comparison, we use metal abundances  relative to hydrogen  
(X/H).

\section{Comparison with the ionized gas}\label{sec:compa}

\begin{figure}[t!] 
\hspace{0.2cm} 
\epsfxsize=5.5cm 
\rotatebox{-90}{\epsfbox{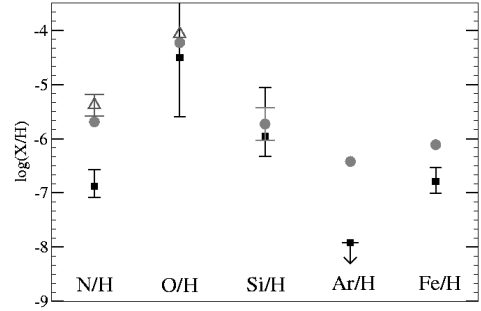}} 
      \caption[]{\small{Comparison of X/H between H\1\ and H\2\ regions  in \IZ. Error bars are at~2$\sigma$. Squares:~this study (assuming the abundances of Sect.~\ref{sec:abundances} and using Ar\1\ for argon i.e. without any ionization correction). Triangles:~V83. Circles:~Izotov~et~al.~(1997) for N, O, Ar, and Fe. The silicon abundance in the ionized region, not measured in \IZ\ is assumed to be equal to $-$1.5$\pm$0.1 which is the mean value observed in BCDs (Izotov~\& Thuan~1999).}} 
         \label{fig:compa1} 
\vspace{0.1cm} 
\end{figure}

Figure~\ref{fig:compa1} compares several abundances ratios  representative of 
  the ionized phase of \IZ,  derived from optical and UV emission lines,   
with those of the neutral gas. 
 We define the underabundance  $\delta_{\textrm{\scriptsize{H}\tiny{I}}}$(\textit{x}), as the  
logarithmic difference of an abundance ratio \textit{x} between  the H\2\ and the H\1\ regions by~:\\ 
 
$\delta_{\textrm{\scriptsize{H}\tiny{I}}}~(x)=log~(x)_{\textrm{\scriptsize{H}\tiny{II}}}-log~(x)_{\textrm{\scriptsize{H}\tiny{I}}}$\\ 
 
The column densities of  oxygen and silicon of \IZ\ are similar in the neutral  
and in the ionized gas within large uncertainties. However the  result 
for nitrogen is much more reliable, and is certainly the best established 
result of this study, yielding  
$\delta_{\textrm{\scriptsize{H}\tiny{I}}}$(N/H)=1.2$\pm$0.3; in other words 
 N is about fifteen times less  abundant  in the neutral gas with respect to the ionized gas. This difference is larger than  3$\sigma$.  
Other metals for which abundances are rather 
 accurately determined also show  underabundances:~$\delta_{\textrm{\scriptsize{H}\tiny{I}}}$(Fe/H)=0.7$\pm$0.2 and  
$\delta_{\textrm{\scriptsize{H}\tiny{I}}}$(Ar/H)$>$1.5$\pm$0.1 (without any attempt to correct for partly--ionized, low--density regions).

The following discussion aims to identify the 
effects responsible for the  observed iron, nitrogen, argon, and oxygen underabundances.

\subsection{Iron and silicon}\label{sec:iron} 
A concern with the abundance of iron in H\2\ regions is that it is derived from the  weak  $\lambda$4658 line of Fe\3, and  is  relatively uncertain. In  
any case, we find that iron is underabundant by  
$\delta_{\textrm{\scriptsize{H}\tiny{I}}}$(Fe/H)=0.7$\pm$0.2 in \IZ. 
 This underabundance might be overestimated since it is likely that iron is more efficiently depleted on dust grains in the  
neutral phase than in the ionized phase.

In \IZ, the only one available Si\2\ line is slightly saturated (Sect.~\ref{sec:CDothers}) so that the determination of the Si\2\ column density  is  uncertain. Within the large error bars, the abundances in the ionized region and in the neutral region are in  agreement.

Because of all these uncertainties, and because silicon and iron  can be depleted on dust grains, we prefer to focus on nitrogen abundance which is much more reliable.

\subsection{Nitrogen}\label{sec:nitrogen} 

\begin{figure}[t!] 
\hspace{0.2cm} 
\epsfxsize=8cm 
\rotatebox{180}{\epsfbox{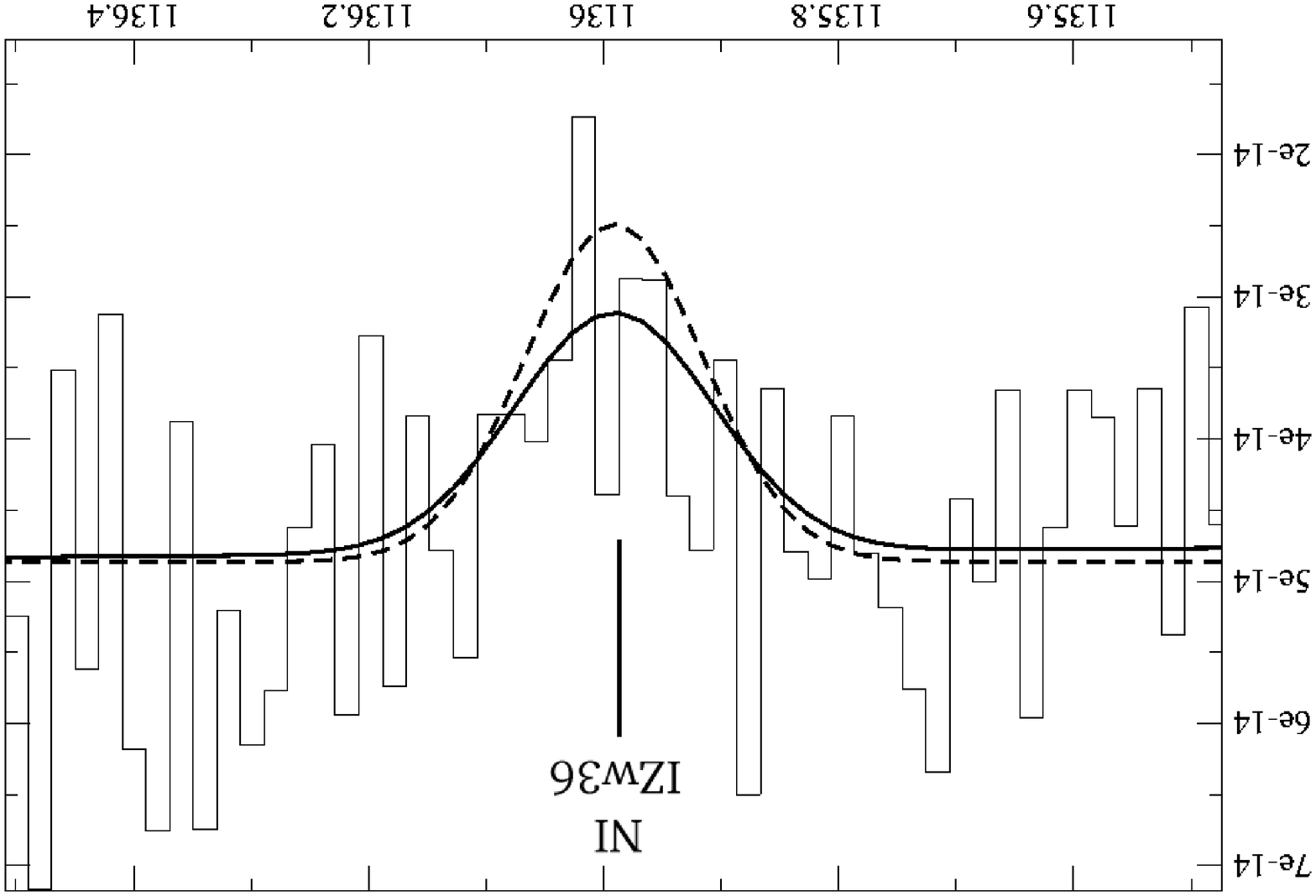}} 
      \caption[]{\small{Fit of the $\lambda$1134.98 line.  Data are binned by a factor 2 for  display purposes. The solid line is for log~(N\1/H\1)=$-$6.88 and shows our best fit. The dashed line is for log~(N\1/H\1)=$-$5.63  and does not fit well the data.}} 
         \label{fig:compaNI} 
\vspace{0.1cm} 
\end{figure}

The underabundance in \IZ\ $\delta_{\textrm{\scriptsize{H}\tiny{I}}}$(N/H)=1.2$\pm$0.3 is  too large to be   
accounted for by chemical inhomogeneities alone in the  neutral gas,  
given the underabundance of iron.
In order to illustrate the underabundance of nitrogen in the neutral gas, we have compared in Fig.~\ref{fig:compaNI} our best fit of the $\lambda$1134.98 N\1\ line, one of the 6 lines we used, with the profile assuming the N/H ratio as in the ionized gas. The two profiles are inconsistent within 3$\sigma$.

Furthermore, even including the largest possible ionization correction (i.e. including N\2\ and taking the upper error bars on N\1\ and N\2\ column densities), the underabundance of nitrogen is still positive: $\delta_{\textrm{\scriptsize{H}\tiny{I}}}$(N/H)=1.0$^{+0.2}_{-0.9}$. This 
supposes that 75\%\ of the nitrogen is ionized into N\2\ in the neutral gas of \IZ. Although the determination of N\2\ column density is uncertain (Sect.~\ref{sec:CDni}), we can reasonnably conclude that nitrogen is indeed underabundant in the H\1\ gas of \IZ.

\subsection{Argon}\label{sec:argon} 
The  argon deficiency is  $\delta_{\textrm{\scriptsize{H}\tiny{I}}}$(Ar/H)$>$1.5$\pm$0.1. 
Argon is produced in massive stars and should not be depleted in a  
 low--density H\1\ cloud (Jenkins~et~al.~2000). 
A major part of the underabundance in the neutral gas is certainly due to an ionization correction from low--density, partly--ionized regions of 
the ISM (Sect.~\ref{sec:abundances}) such as argon is mainly ionized into Ar\2.

\subsection{Oxygen}\label{sec:oxygen} 
 
The O/H ratio in the neutral region of \IZ\ is difficult to assess because  
of the large errors on the O\1\ column density (Table~\ref{tab:CD}). However,  
it is possible to  estimate its value using other elements which are produced in the same massive stars.
 Phosphorus turns out to be a good tracer of oxygen (Lebouteiller~2003), since argon  
is partly--ionized in the neutral region (Sect.~\ref{sec:argon}) and since  
the depletion of silicon on dust grains is uncertain. 
Depletion is not a concern for phosphorus since it is not much   
depleted, and in the same way  as oxygen (Savage~\&~Sembach~1996).

  We calculate O$^P$ which estimates  the neutral oxygen abundance assuming 
$[$P/O$]$$\approx$0 in  
the neutral gas (using the recommended solar value of Lodders~(2003) log~(P/O)$_\odot$=$-$3.23$\pm$0.09). 
We find  log~(O$^P$/H)$\approx$$-$5.1$\pm$0.5, giving an underabundance  
$\delta_{\textrm{\scriptsize{H}\tiny{I}}}$(O$^P$/H)$\approx$0.9$\pm$0.5.
 Within the errors, this estimation of the  oxygen abundance
is lower  with respect to that in the ionized gas. 
Given the low depletion of phosphorus and oxygen in the Galactic ISM (Andr\'e~et~al.~2003 and Savage~\&~Sembach~1996), it is very likely that oxygen is genuinely
deficient in the neutral medium of \IZ.

Assuming log~(O$^P$/H)=$-$5.1$\pm$0.5 in the H\1\ gas, we find
 log~(N/O$^P$)=$-$1.8$\pm$0.6 (2$\sigma$ uncertainty) lower than, although consistent with,  the value  $-$1.49$\pm$0.01 in the ionized gas (Izotov~\&~Thuan~1999).
Also, we find  log~(Ar/O$^P$)$<$$-$3.0 as compared to the solar value log~(Ar/O)$_\odot$=$-$2.1$\pm$0.1 (Lodders~2003), confirming that argon is likely ionized into Ar\2\ as stressed in Sect.~\ref{sec:argon}.

\section{Conclusions}\label{sec:cncl}

Several heavy  elements in the neutral gas of \IZ\  
are underabundant in the neutral gas with respect to the ionized gas. 
Various effects can be responsible for the underabundances  
$\delta_{\textrm{\scriptsize{H}\tiny{I}}}$(X/H):\\ 
$-$ The presence of an unprocessed neutral gas, less chemically evolved, in the line of sight  
can reduce all the abundances X/H in the neutral gas with respect to the ionized gas.\\ 
$-$ A more efficient depletion on dust grains in the neutral medium can account for the iron deficiency.\\ 
$-$ An additional metal enrichment could be responsible for nitrogen and oxygen (using phosphorus as a tracer) overabundance in the ionized gas, although in this case, silicon which is an $\alpha$-element should be also overabundant (but see caption of Fig.\ref{fig:compa1}).

 Our findings that  metals  are observed in the
neutral region of \IZ\ indicates that this gas phase has been already enriched by previous star formation, either quiescent or in bursts (Legrand~2000). 
Schulte--Ladbeck~et~al.~(2001) detect indeed an old stellar population which could have  enriched the neutral region of \IZ.

In this study, we considered H\2\ region abundances derived from forbidden lines in the visible and the ultraviolet. 
However, the abundances derived in this way might  be underestimated by factors ($\gtrsim$2 for oxygen for instance) as claimed by Tsamis~et~al.~(2003) compared to the more reliable 
 abundances deduced from optical  recombination lines or  
far-IR forbidden lines (which unfortunately are not  
available for the BCDs). Consequently, the  abundances differences between the two gas phases of \IZ\ could be even larger than those given in Fig.~\ref{fig:compa1}.

At present time,  the overall picture still remains unclear. 
To assess which of the  effects (unprocessed neutral gas, depletion  
on dust grains, metal enrichment) are dominant, it is necessary to 
 investigate the neutral gas of more gas--rich star--forming galaxies.  
Furthermore, with a larger sample, we can  
reasonnably expect to find several good oxygen determinations.

\begin{acknowledgements} 
This work is based on data obtained for the French Guaranteed  
  Time and the PI Team Guaranteed Time by the NASA-CNES-CSA \textit{FUSE}  
  mission operated by the Johns Hopkins University. French  
  participants are supported by CNES. 
     This work has been done using the profile fitting procedure \texttt{Owens.f} developed by M. Lemoine and the \textit{FUSE} French Team. We are very grateful to Martial Andr\'e for providing useful comments. 
\end{acknowledgements}

\end{document}